\documentclass{PoS}
\usepackage{overpic}
\usepackage{amsmath}
\pdfoutput=1
\title{Long-range correlation studies at the SPS energies in MC model with string fusion}

\ShortTitle{Long-range correlation studies at the SPS energies in MC model with string fusion}

\author{\speaker{Vladimir Kovalenko}
                                         \\
        Saint Petersburg State University\\
        E-mail: \email{nvkinf@rambler.ru}}

\author{Vladimir Vechernin\\
        Saint Petersburg State University\\
        E-mail: \email{vechernin@gmail.com}}

\abstract{Studies of the ultrarelativistic collisions of hadrons and nuclei at different centrality and energy
enable to explore the QCD phase diagram in a wide range of
temperature and baryon density. Long-range correlation studies are considered as a tool, sensitive to the observation of phase transition and the critical point.
In the present work, a Monte Carlo model of
proton-proton, proton-nucleus, and nucleus-nucleus collisions is applied to heavy
and light ion collisions at the cms energy range from a few up to several hundred GeV
per nucleon. The model describes the nuclear collisions at the
partonic level through interaction of color dipoles and takes into account the
effects of string fusion, which can be considered as an alternative to relativistic hydrodynamics way of describing the collective phenomena in heavy-ion collisions. The implementing
of both the string fusion and the finite rapidity length of strings allowed to consider the
particle production at non-zero baryochemical potential. We calculated the long-range
correlation functions and correlation coefficients between multiplicities and transverse momentum at several energies for different colliding systems and obtained predictions for the experiment.

}

\FullConference{XXII International Baldin Seminar on High Energy Physics Problems,\\
		15-20 September 2014\\
		JINR, Dubna, Russia}

\begin{document}

\section{Introduction}
The exploration of the phase diagram of strongly interacting matter and a search for the onset of deconfinement and the critical point is one of the main subjects of heavy ion physics.  There is a general consensus that at zero baryochemical potential the deconfinement phase transition is a smooth cross-over \cite{PD2}, which is mainly based on the lattice QCD calculations. However, the numerical lattice study of the behaviour of the QCD matter at non-zero baryon density, due to the sign problem, is not straightforward \cite{HW,HW1,HW2}. According to existing calculations \cite{PD0,PD1,PD3}
it is expected that at high net baryon density, this transition is of the type of the first order, which suggests an existence of a critical point in the phase diagram at the intermediate net-baryon density. However, in several works a smooth crossover has been obtained at any value of baryochemical potential \cite{HWn,HWm}.

 The experimental investigations on the QCD phase diagram are related to the study of the collisions of the ions at high energy \cite{PDE0,PDE1}. The study of the QCD phase diagram is a part of physical program of the NA61 experiment at
the SPS, experiments at RHIC and also future detectors CBM at FAIR and MPD at NICA
 \cite{BES0,BES1,BES2,BES3}.
Among the observables, being used in these studies, the most sensitive ones, such as
a collective flow, correlations, fluctuations, require event-by-event analysis \cite{PDO0, NA61}. Particularly, the studies of long-range correlations
between variables taken from two different rapidity windows are included in the research program of the experiment NA61 as a tool, sensitive to the observation of phase transition and the critical point \cite{NA61}.

For the correct interpretation of experimental findings and evaluation of the sensitivity  
of experimental methods, 
the theoretical modelling of the evolution of a heavy-ion collision with the explicit calculation
of observables in the conditions, close to the experimental ones, is required.
Due to the complexity of the evolution process of a heavy-ion collision and inapplicability
of QCD perturbation theory in the low momentum
region, semiphenomenological models are widely applied in this area.

One of the models, which is used 
to describe the properties of the initial dense state of the strongly interacting
matter that occurs immediately after the nuclear interaction of high energy, is the model for the formation
and fusion of quark-gluon strings \cite{SF1,SF2,F33,F33a,OL1}. According to this model, the hadrons,
produced in the soft region, can be described as a result of decay of colour field tubes -- strings, which are
formed between the interacting partons. With increasing energy and the mass number of colliding
nuclei, the density of strings grows, and they begin to overlap forming an area in a transverse plane with
stronger color field ("clusters"). In the limit of high density, it is expected that all cross-section plane will
be a single cluster, which is supposed to be in its physical properties equivalent to the quark-gluon
plasma. Convincing evidence in favour of the fusion of the strings are discovered experimentally growth
of the mean transverse momentum with multiplicity in hadron and nuclear collisions and a significant decrease in the yield of
multiplicity in the collisions of heavy ions compared to models with independent strings, which is compatible with the existing experimental data on multiplicity. It is assumed
that fusion and percolation of strings responsible for the appearance of the ridge in two-particle
correlations \cite{TP1,TP2}.
In the recent papers \cite{RP1,RP2,RP3}
it was shown that the equation of state of QGP at zero chemical potential obtained in the colour string
percolation model is in an excellent agreement with the lattice results. In addition, the approach has been successfully applied for the determination of the shear viscosity \cite{RP4} of QCD matter, with was found in a good agreement with experimental data in a wide range of temperatures.

 In the framework of the string fusion approach, the critical behaviour is expected when the processes of string fusion and percolation come into play, what can be considered
as a possible way of quark-gluon plasma formation. Around percolation threshold, strong
fluctuations in colors of strings appear what lead to large fluctuations in some observables, which
one can find by the event by event analysis \cite{NA61}.
In the present work, the model is applied to heavy and light ion collisions at the center of mass energy range from a few up to several hundred GeV per nucleon. In the next sections, we very briefly describe the main features of the Monte Carlo model with string fusion and explain how it can be applied at non-zero baryon density. Then, we introduce the observables (correlation functions and correlation coefficients) used in the study. Finally, we present the results and give our conclusions.

\section{The Monte Carlo model}
The present model \cite{OL2, YF1PoS2} describes nuclear collisions on the partonic level, without referring to the Glauber picture of the independent collisions of the nucleons.
Initially, the nucleons are supposed to be distributed in a nucleus according to
the charge density distribution $\rho(r)$. We used harmonic oscillator model for $^7$Be and $^9$Be nuclei and Woods Saxon distribution for heavier species~\cite{WoSa}.

\renewcommand{\arraystretch}{2.4}
\begin{table}[h]
\begin{tabular}{|c|c|c|c|}
\hline
  \textbf{Nucleus}
  & \textbf{Model}
  & \textbf{$\rho(r)$}
  &  \textbf{Parameters}  \\
\hline
$^{208}$Pb & Woods Saxon & $\dfrac{\rho_0}{1+\exp[(r-R)/d]}$ & $R=6.63$ fm, $d=0.545$ fm   \\
$^{197}$Au & Woods Saxon & $\dfrac{\rho_0}{1+\exp[(r-R)/d]}$ & $R=6.38$ fm, $d=0.535$ fm   \\
$^{40}$Ca, $^{40}$Ar  & Woods Saxon & $\dfrac{\rho_0}{1+\exp[(r-R)/d]}$ & $R=3.53$ fm, $d=0.542$ fm  \\
$^{7}$Be & Harmonic oscillator & $\rho_0 \left(1+a(r/r_0)^2\right)  \exp \left(-\left({r/r_0}\right)^2\right)$ & $r_0$=1.77 fm, a=0.327  \\
$^{9}$Be & Harmonic oscillator &  $\rho_0  \left(1+a(r/r_0)^2\right)  \exp \left(-\left({r/r_0}\right)^2\right)$ & $r_0$=1.791 fm, a=0.611  \\
\hline
\end{tabular}
\caption{Nuclear distributions and parameters used in the present study.}
\label{Table1}
\end{table}

In the model, each nucleon contains a valence quark-diquark pair and several sea quark-antiquark pairs. The number of sea pairs is generated according to Poisson distribution around some mean value, which is energy-dependent and is adjusted to describe the total inelastic cross section in pp collisions. The transverse coordinates of these partons are distributed around the center of each nucleon according to two-dimensional Gauss distribution with a mean-square radius $r_0$.  
It is assumed that quark-diquark and quark-antiquark pairs form dipoles.
An elementary interaction according to the color dipole approach \cite{OL7, OL8} is realized with the  
probability amplitude given by:
\begin{equation} \label{newformula}
			f=\frac{\alpha_S^2}{2}\Big[ K_0\left(\frac{|\vec{r}_1-\vec{r}_1'|}{r_{\mathrm{max}}}\right) +
			K_0\left(\frac{|\vec{r}_2-\vec{r}_2'|}{r_{\mathrm{max}}}\right) 
			- K_0\left(\frac{|\vec{r}_1-\vec{r}_2'|}{r_{\mathrm{max}}}\right)
			- K_0\left(\frac{|\vec{r}_2-\vec{r}_1'|}{r_{\mathrm{max}}}\right)	\Big]^2,
\end{equation}	
where $K_0$ is modified Bessel function of the second order, $(\vec{r}_1, \vec{r}_2), (\vec{r}_1', \vec{r}_2')$ are transverse coordinates of the projectile and target dipoles, and $\alpha_s$ --  effective coupling constant,
$r_{\mathrm{max}}$ is characteristic confinement scale. According to this formula, two dipoles interact with higher probability, if the ends of dipoles are close to each other in the transverse plane, and (other equal) if they are wide.

In our Monte Carlo model it is assumed, that if there is a collision between two dipoles,
 two quark-gluon strings
are stretched between the ends of the dipoles,
and the process of string fragmentation gives
observable particles. The particle production of a string is assumed to go  uniformly between the string rapidity ends $y_{\mathrm{min}}$ and $y_{\mathrm{max}}$ \cite{OL2,OL6}, which are related to the longitudinal momentum fraction, carried by the parons forming the string. The emission of the charged particles is generated with mean number per rapidity $\langle \mu \rangle$, independently in each rapidity interval, with Poisson distribution.

The transverse position of string is assigned to the arithmetic mean
of the transverse coordinates of the partons at the ends of the string.
Due to finite transverse size of the strings they overlap, that in the framework of string fusion model
 \cite{SF1, SF2} gives the source of a higher tension.
We used  the lattice variant of the model \cite{VK1, VK2}, according to which strings are
supposed to be fused if their centres occupy the same cell. The area of each cell is equal to the transverse string
area ($S_{\mathrm{str}}=\pi r^2_{\mathrm{str}}$). In the calculations, we also assume local string fusion scenario \cite{VK1, VK2}. Each bin in rapidity with an integer number of overlapped strings is processed separately. Mean multiplicity of charged particles and mean $p_t$ originated from the cell where $k$ strings are overlapping are obtained according the following:
\begin{equation} \label{muptloc}\nonumber
	\left\langle \mu\right\rangle_k=\mu_0 \sqrt{k}, \hspace*{1cm}
	\left\langle p_t\right\rangle_k=p_0 \sqrt[4]{k}.
\end{equation}	 
Here $\mu_0$ and $p_0$ are mean charged multiplicity from one single string per rapidity unit
and mean transverse momentum from one single string. 
The relations allows to calculate long-range correlation functions and
correlation coefficients between
multiplicities and mean transverse momentum of charged particles \cite{YF1PoS2}.
Important, that $\mu_0$ and $p_0$ cancels out in the correlation observables defined below and do not influence the final result on correlation coefficients,
which makes the calculations more robust. 

The parameters of the model $\alpha_s, r_0, r_{\mathrm{max}}, \mu_0$ are supposed to be energy-independent and have been constrained from charged particles multiplicity in a wide energy range (from ISR to LHC) in pp collisions and in minimum bias p-Pb and Pb-Pb collisions at the LHC energy. The detailed procedure of parameters tuning is described in \cite{OL6}.

\vspace{0.5cm}

Important to note, that the implementation of the string fusion in the model, described above, allows us to
separate phenomena taking place (in the same event) in different rapidity regions.
At the SPS energies, the contribution of sea quarks and corresponding strings is not significant. Therefore, the valence strings dominate. Due to the fact that the diquark carries out larger longitudinal momentum fraction \cite{OL2}, the valence string, formed by a quark-diquark pair, is asymmetric. Most of the particles in
forward rapidity region are coming from valence strings
from their diquark parts.

\begin{figure}[h]
\hspace{1cm}  \includegraphics[height=.22\textheight]{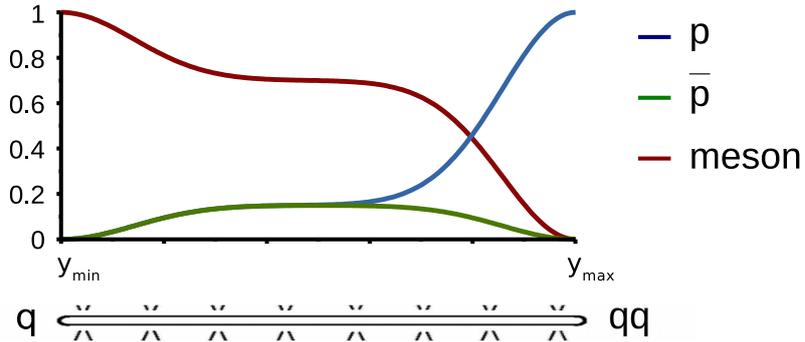}
  \caption{A sketch of a particle composition, coming from one string: fractions of protons, antiprotons and mesons as a function of rapidity.
\newline
\ }
  \label{fig1}
\end{figure}

\begin{figure}[h]\centering
\includegraphics[height=.33\textheight]{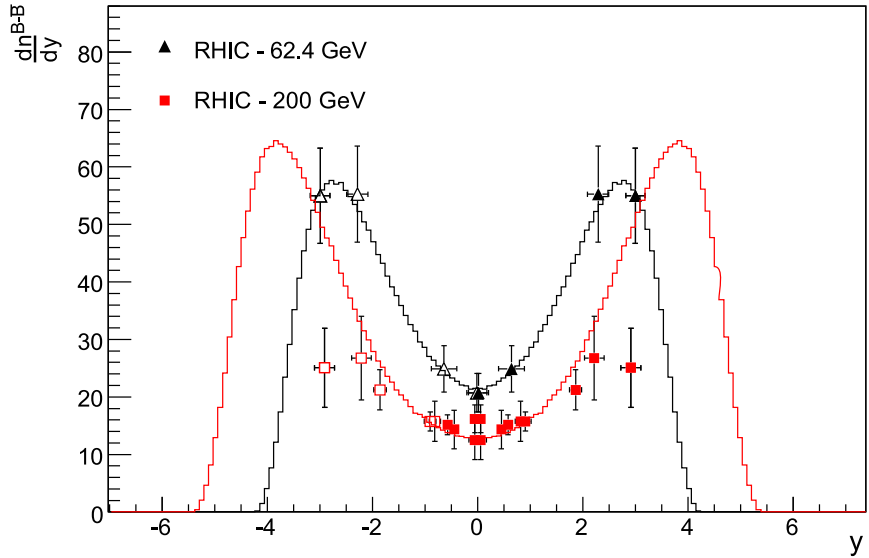}
  \caption{Net baryon rapidity distribution in Au+Au collisions. Points -- experimental data at RHIC, lines -- results of calculations in a valence string model \cite{VSM}.}
  \label{fig1a}
\end{figure}

The approximate particle content of the valence rapidity string is shown in Fig. \ref{fig1}. The diquark end of the string, which is generally located in the 
forward rapidity region, is characterized by large 
baryon yield. In contrary, the opposite end, which carries out a quark, emits mostly pions, which happens
around midrapidity. Altogether, it provides a distribution of a net-baryon density over rapidity with large baryon redundancy in the forward region, which has been obtained in the experimental data (Fig. \ref{fig1a}). We note, that the valence string model, developed in \cite{VSM}, successfully describes net baryon yields in a wide energy range.
\vspace{1cm}

\vspace{0.6cm}

\section{Long-range correlations}

The long-range correlations between observables in two separated rapidity windows (backward $B$ and forward $F$) are numerically characterized by correlation functions and correlation coefficients. 
The correlation function is defined as an average of the observable on backward window at fixed value of the observable in the forward window:

\begin{equation}
f(F)={\langle B \rangle}_{F}.
\end{equation}

Both $B$ and $F$ could be either $N_{\mathrm{ch}}$ -- the number of charged particles in the rapidity window or $p_t$ -- mean in the event transverse
momentum of charged particles in the given window:
\begin{equation}
p_t = \dfrac{1}{N_{\mathrm{ch}}} \sum\limits_{i=1}^{N_{\mathrm{ch}}}{{p_t}_i}.
\end{equation}

Correspondingly, tree types of correlation functions are defined \cite{VK1, VK2}: \ n-n ($f_{\mathrm{n-n}}$),\newline pt-n ($f_{\mathrm{p_t-n}}$) and pt-pt ($f_{\mathrm{p_t-p_t}}$).
\begin{equation}
f_{\mathrm{n-n}}(n_F)={\langle n_B \rangle}_{n_F}, \hspace{1cm}
f_{\mathrm{p_t-n}}(n_F)={\langle {p_t}_B \rangle}_{n_F}, \hspace{1cm}
f_{\mathrm{p_t-p_t}}(n_F)={\langle {p_t}_B \rangle}_{{p_t}_F}.
\end{equation}

The correlation coefficient represents the slope
of the correlation function:
\begin{equation} \label{defcorr}
b_{B-F}=\frac{df(F)}{dF}|_{{F}=<{F}>}.
\end{equation}

In order to reduce experimental and theoretical uncertainties, it is often useful to switch to
normalized variables: $B \rightarrow B/{\langle B\rangle}, F \rightarrow F/{\langle F\rangle}$. In this case, the
 $p_t-n$ correlation coefficient becomes dimensionless,
 and also both $n-n$ and $p_t-p_t$ correlation coefficients do not change in case of symmetrical rapidity windows.
 In normalized variables:

\begin{eqnarray}
b_{nn}^{}&=&\frac{<n_F>}{<n_B>} \cdot\frac{d<{n}_B>}{dn_F}|_{n_{F}=<n_{F}>}, \\
b_{p_t-n}^{}&=&\frac{<{n}_F>}{<{p_t}_B>} \cdot\frac{d<{p_t}_B>} {d{n}_F}|_{n_{F}=<n_{F}>}, \\
b_{p_t-p_t}^{}&=&\frac{<{p_t}_F>}{<{p_t}_B>} \cdot\frac{d<{{p_t}}_B>}{d{p_t}_F}|_{{p_t}_{F}=<{p_t}_{F}>}.
\end{eqnarray}

\scalebox{.956}[1.0]{There is an alternative definition on long-range correlation coefficient} \cite{Adam:2015mya}: 
$b_{\mathrm{corr}}=\frac{\langle n_B n_F \rangle - \langle n_B \rangle \langle n_F \rangle}{\langle n_F^2 \rangle - \langle n_F \rangle^2}$. The definitions are equivalent in case of linear correlation functions, and both of them are used in theoretical and experimental studies \cite{TbD0,TbD1,TbD2,TbD3}.

In our calculations we took p+p,  $^{7}$Be+$^{9}$Be,  p+Pb,  Ar+Ca,  Au+Au and  Pb+Pb collisions at  
the colliding energies $\sqrt{s}$ from 5 GeV to  62.4 GeV. The set of the colliding systems and energy range have been selected in accordance with the experimental program of NA61 Collaboration, BESII program at RHIC and future detectors at FAIR and NICA facilities.

For p+p, Be+Be, p+Pb collisions minimum
bias events have been taken; for heavier species
(Ar+Ca, Au+Au, Pb+Pb) 
two centrality classes have been considered: the central events, which correspond to the number of participant nucleons $N_{\mathrm{part}}>(A+B)/2$,  and the peripheral ones, with $N_{\mathrm{part}} \le (A+B)/2$.
Here, $A$ and $B$ are mass numbers of the colliding nuclei.

\pagebreak 
We selected three rapidity windows configurations:\vspace{-0.3cm}
\begin{center}
(-1 ; 0) -- (0 ; 1); \\
( 0 ; 1) -- (1 ; 2); \\ 
( 0 ; 1) -- (2 ; 3). \\
\end{center}\vspace{-0.3cm}
These windows correspond to mid-rapidity configuration (first two) and middle-forward case (third one). The forward window in the third configuration probes the rapidity region, where significant net baryon density is expected (see Fig. \ref{fig1a}).
\section{Results}
\subsection{Correlation functions}
\begin{figure}[h]
\centering
  \includegraphics[width=0.97\textwidth]{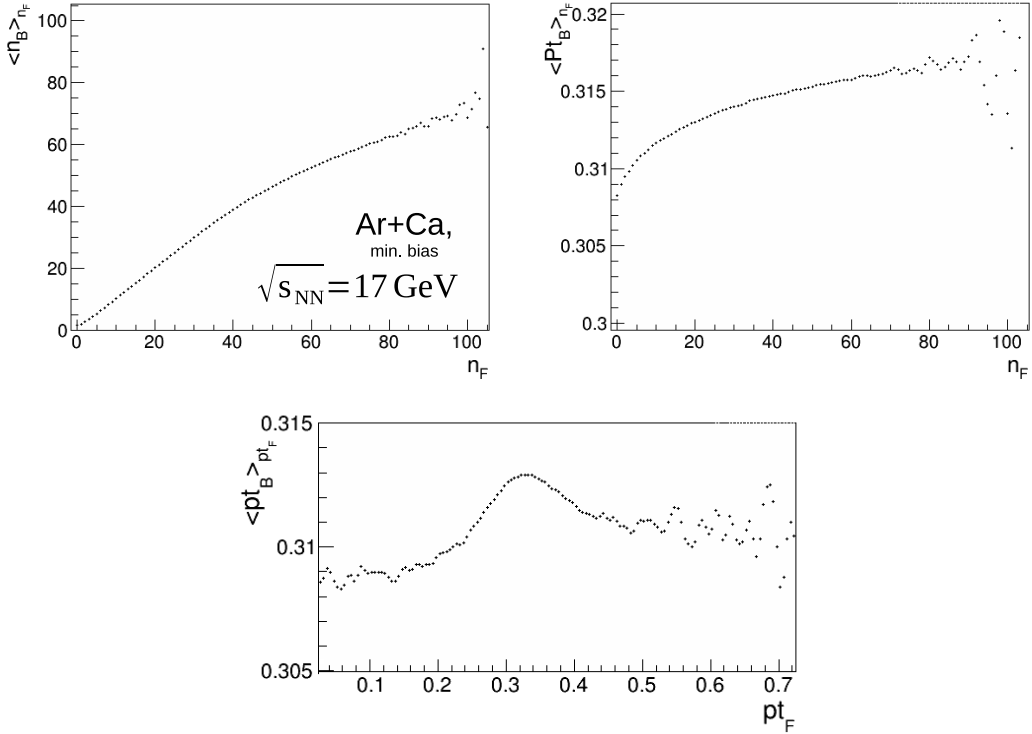}
  \vspace{-0.2cm}
  \caption{Correlation functions in Ar+CA collisions at $\sqrt{s_{NN}}$=17GeV: n-n correlation  (left), pt-n correlation (right) and  pt-pt correlation (bottom). Rapidity windows configuration is (-1;0), (0;1).}
  \label{fig2}
\end{figure}

Fig. \ref{fig2} shows all three types of correlation functions. The n-n correlation function is obtained to be close to linear at low values of $n_F$, which correspond to 
peripheral collisions. The slope of the correlation function decreases with $n_F$ and tends to saturate at high values of $n_F$.
The results on pt-n correlations demonstrate small positive pt-n slope of pt-n correlation function. The pt-pt correlation function is found strongly non-linear in minimum bias collisions. The predictions are similar to the ones, obtained for Pb+Pb collisions at the SPS energy in string fusion approach
\cite{VK1,VK2} and experimental data of the NA49 experiment \cite{NA49baldin}.

\subsection{Correlation coefficients}

In the Fig. 4 the multiplicity-multiplicity correlation coefficient in proton-proton collisions is plotted as a function of the
center-of-mass collision energy per nucleon $\sqrt{s_{NN}}$. The results demonstrate monotonic increase of the $b_{\mathrm{n-n}}$ with $\sqrt{s}$.

\begin{figure*}[h!]
{ \hspace{1.25cm}
\begin{overpic} [width=0.694\textwidth]{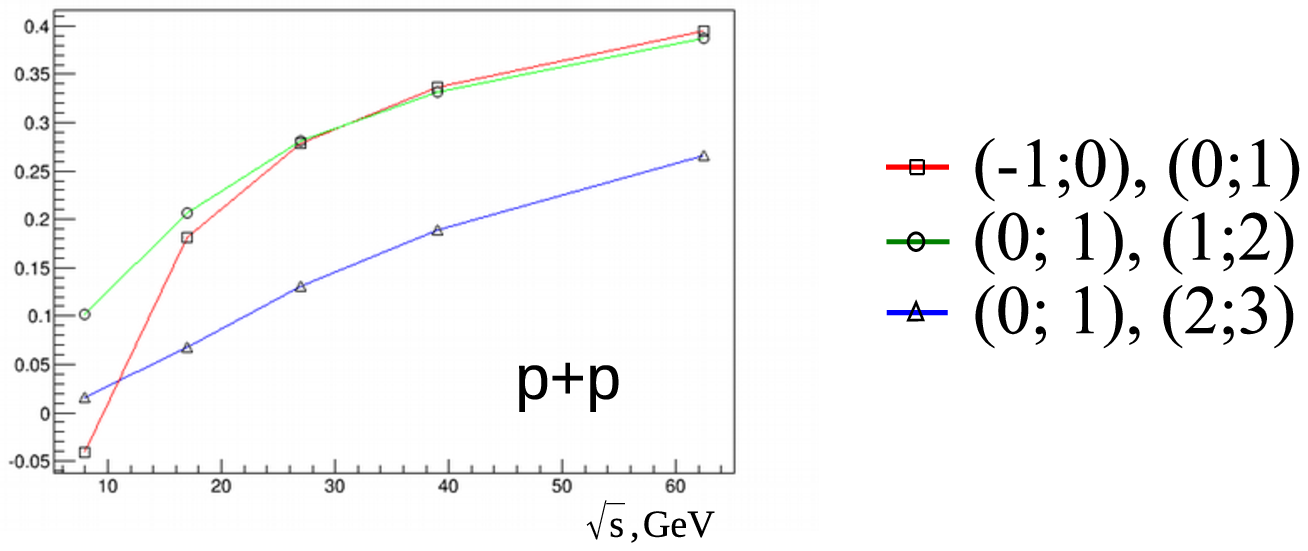}
          \put(7,36){ $\mathrm{b}_{\mathrm{n-n}}$}
\end{overpic}}

\hspace{2cm}
\begin{overpic} [width=0.71\textwidth]{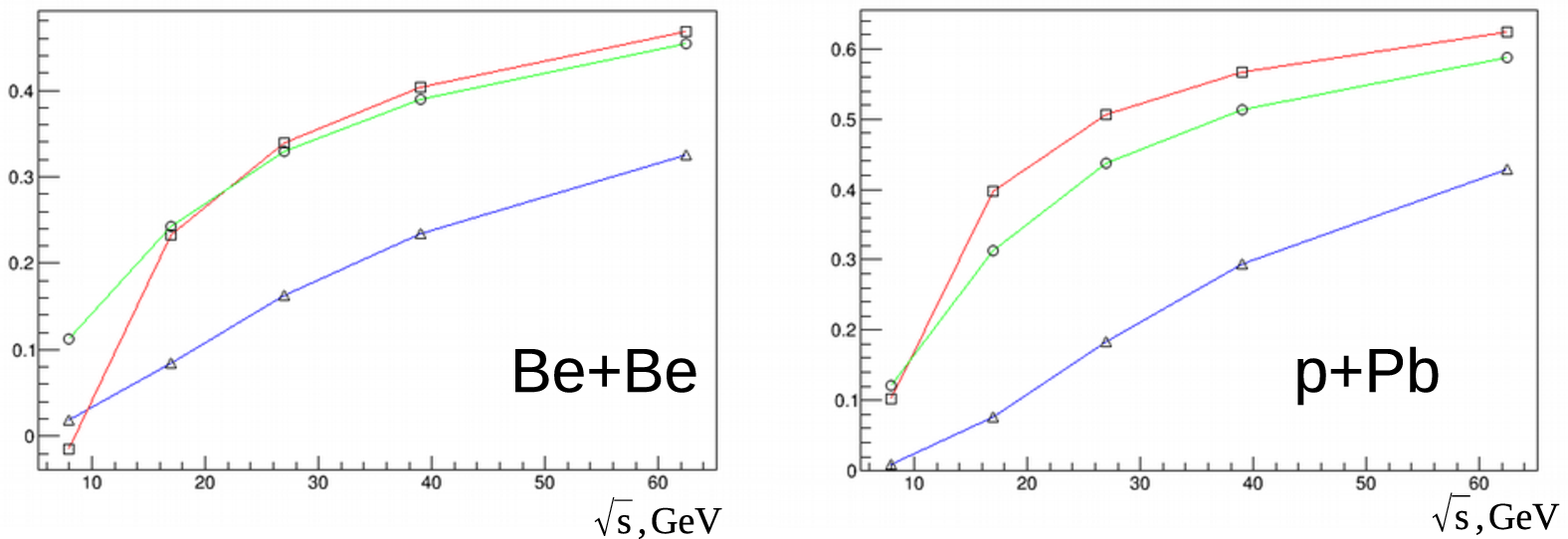}
\end{overpic}
\caption{Dependence of n-n correlation coefficient on the collision energy in proton-proton,  Be+Be and p+Pb collisions.  Three rapidity windows configurations are represented by different colors.}\label{fig4}
\end{figure*}

\begin{figure*}[h!]
\centering
\begin{overpic} [width=0.7\textwidth]{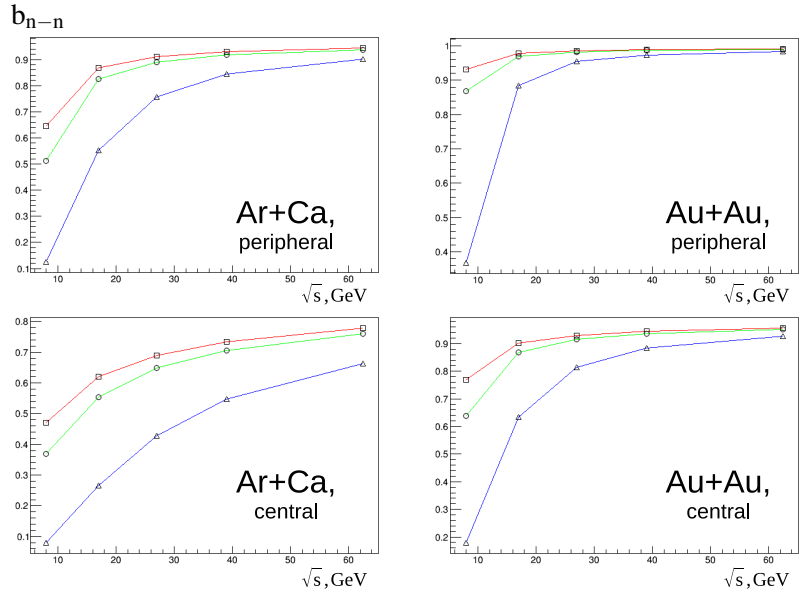}
          \put(0,72){ $\mathrm{b}_{\mathrm{n-n}}$}
\end{overpic}
\caption{Dependence of n-n correlation coefficient in three rapidity windows configurations on the collision energy in Ar+Ca and Au+Au collisions. The color notation is the same as in Fig. 4.}
\label{fig5}
\end{figure*}

\pagebreak

In general, the n-n correlation coefficient is
higher in mid-rapidity and decrease going from 
mid-rapidity to middle-forward configuration. Such behaviour is in agreement with the
previous studies \cite{YF1PoS2,OL15,OL25}, where it
was shown that $b_{\mathrm{n-n}}$ decreases with an increase of the rapidity gap and going to the forward rapidity region.

The results on n-n correlations in proton-ion and light ion collisions are also shown in Fig. \ref{fig4}. The correlation coefficient is significantly higher than the one in pp collisions, but no saturation with the energy is achieved.

Multiplicity-multiplicity correlation coefficients for heavier ion species are plotted in Fig.~\ref{fig5}. The results are split into two different centrality classes, as described above. The Pb+Pb collisions, which have a similar pattern as Au+Au, are not shown. It is found, that the  $b_{\mathrm{n-n}}$ is less in the central class than in the peripheral one. Such behaviour is in a correspondence with the string fusion predictions \cite{OL15}. This fact is also in agreement with a saturation trend of the correlation function (Fig.~\ref{fig2}).
However, one should keep in mind that this observable is subject to the centrality class width dependence \cite{OL15}, and the central class in our calculations
seems to be narrower. So, in order to provide the
more convincing predictions, the calculations in narrow
centrality classes are required.
\vspace{1cm}

\begin{figure*}[h!]
\centering
\begin{overpic} [width=1.05\textwidth]{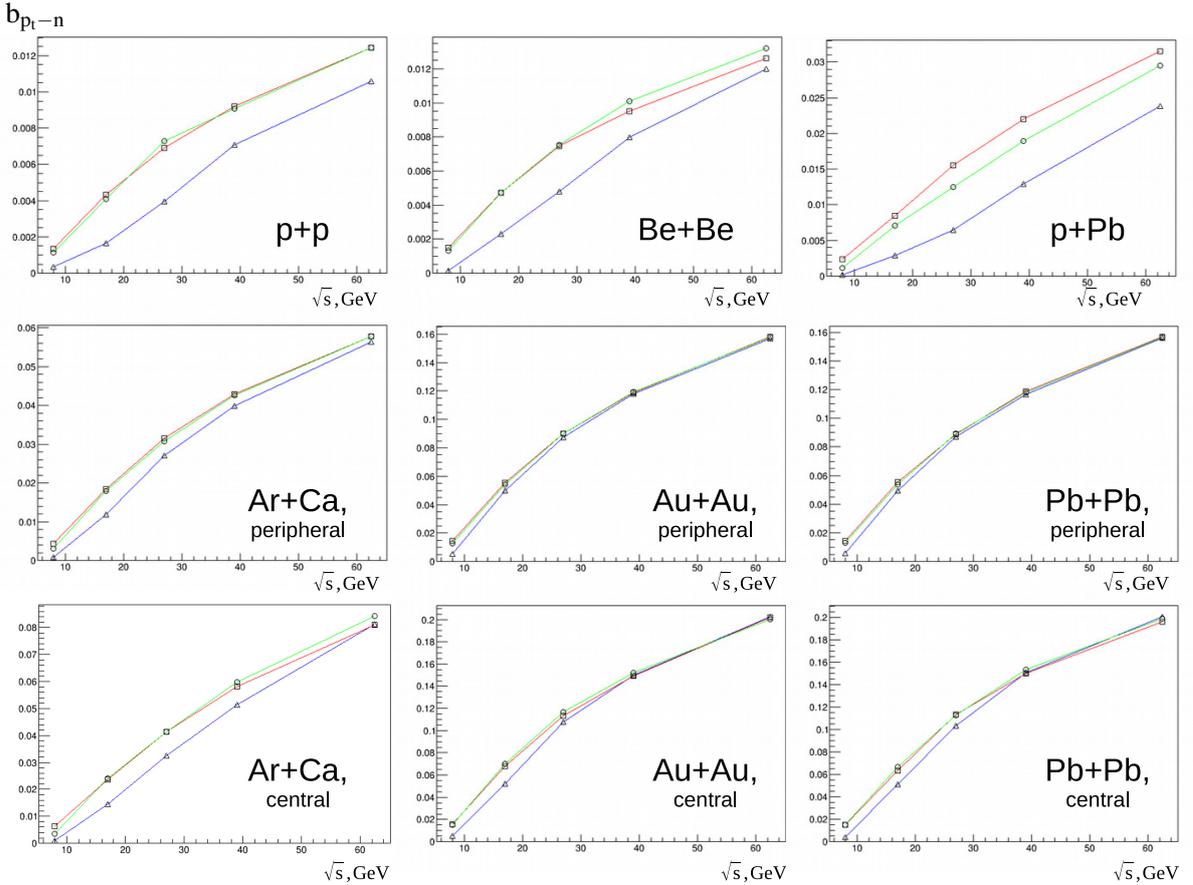}
          \put(0,72){ $\mathrm{b}_{\mathrm{p_t-n}}$}
\end{overpic}
\caption{Dependence of pt-n correlation coefficient in three rapidity windows configurations on the collision energy in various colliding systems. The color notation is the same as in Fig. 4.}
\label{fig6}
\end{figure*}

The results on pt-n correlation coefficient are shown in Fig. \ref{fig6}. We find the smooth monotonic behaviour of $b_{\mathrm{pt-n}}$ with the collision energy, similar to the one, observed for multiplicity 
correlations. We should note, that at minimal energy, the pt-n correlation is practically zero, which is not true for n-n correlations. It may be related to the fact that n-n correlations are present also in the absence of any collectivity, due to the fluctuations
in the number of the emitting sources \cite{VF}. In  contrary, the pt-n correlations appear only in the presence of the string fusion, and their sign depends on the variance of the initial sources \cite{Andronov}. We should note that the magnitude of the pt-n correlations increases significantly from light to heavier nuclei. However, this observable also suffering from the dependency on the centrality selection \cite{OL15}, so a more deep centrality investigation
would improve the calculations.
\vspace{0.6cm}

\begin{figure*}[h!]
\centering
\begin{overpic} [width=1.05\textwidth]{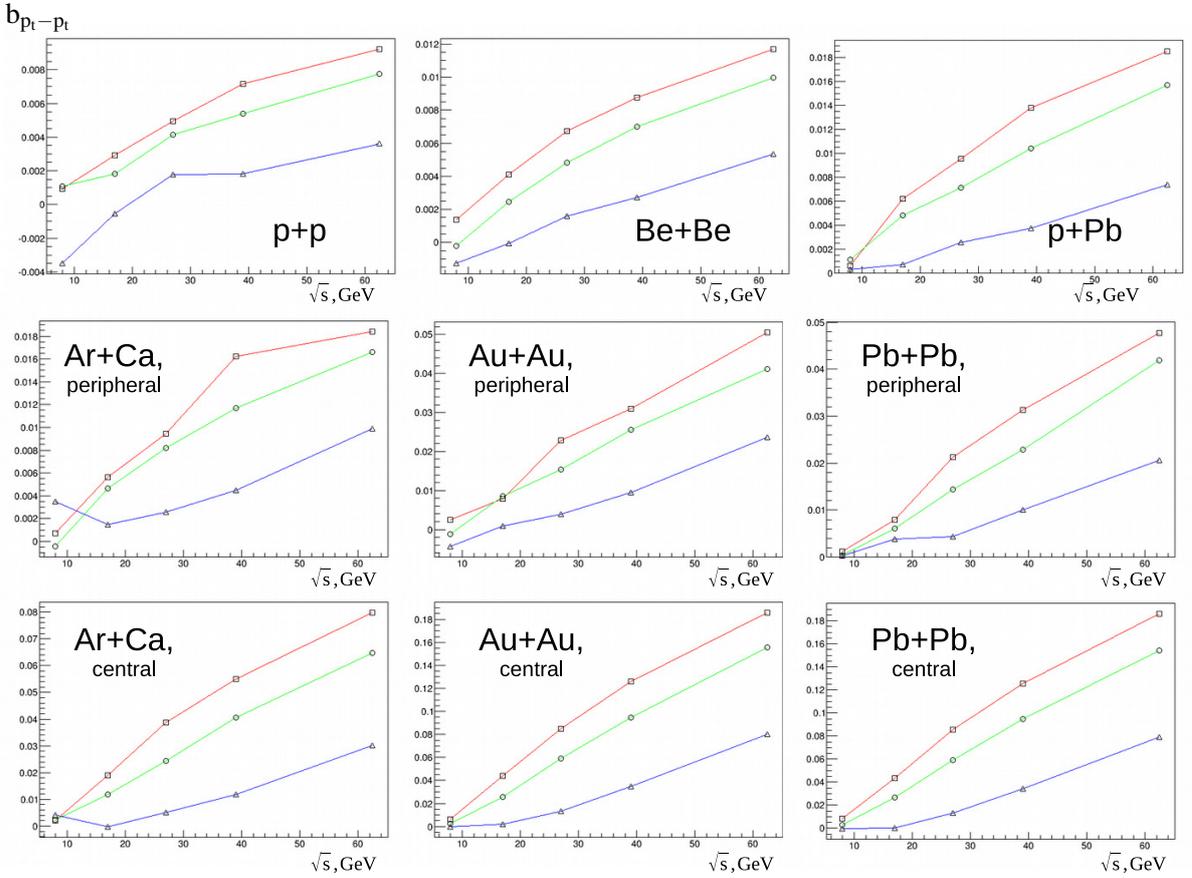}
          \put(0,71){ $\mathrm{b}_{\mathrm{p_t-p_t}}$}
\end{overpic}
\caption{Dependence of pt-pt correlation coefficient in three rapidity windows configurations on the collision energy in various colliding systems. The color notation is the same as in Fig. 4.}
\label{fig7}
\end{figure*}

Fig. \ref{fig7} shows the energy dependence of the transverse momentum
correlation coefficient. Contrary to 
the previously discussed observables,  $b_{\mathrm{pt-pt}}$ is a correlation between two intensive
variables \cite{VK2}, and it is more
robust towards the centrality selection,
because it is less influenced by the
volume  fluctuations due to variation in
participant nucleons number and impact parameter. 

\pagebreak

\begin{figure*}[h!]
\centering
\begin{overpic} [width=0.95\textwidth]{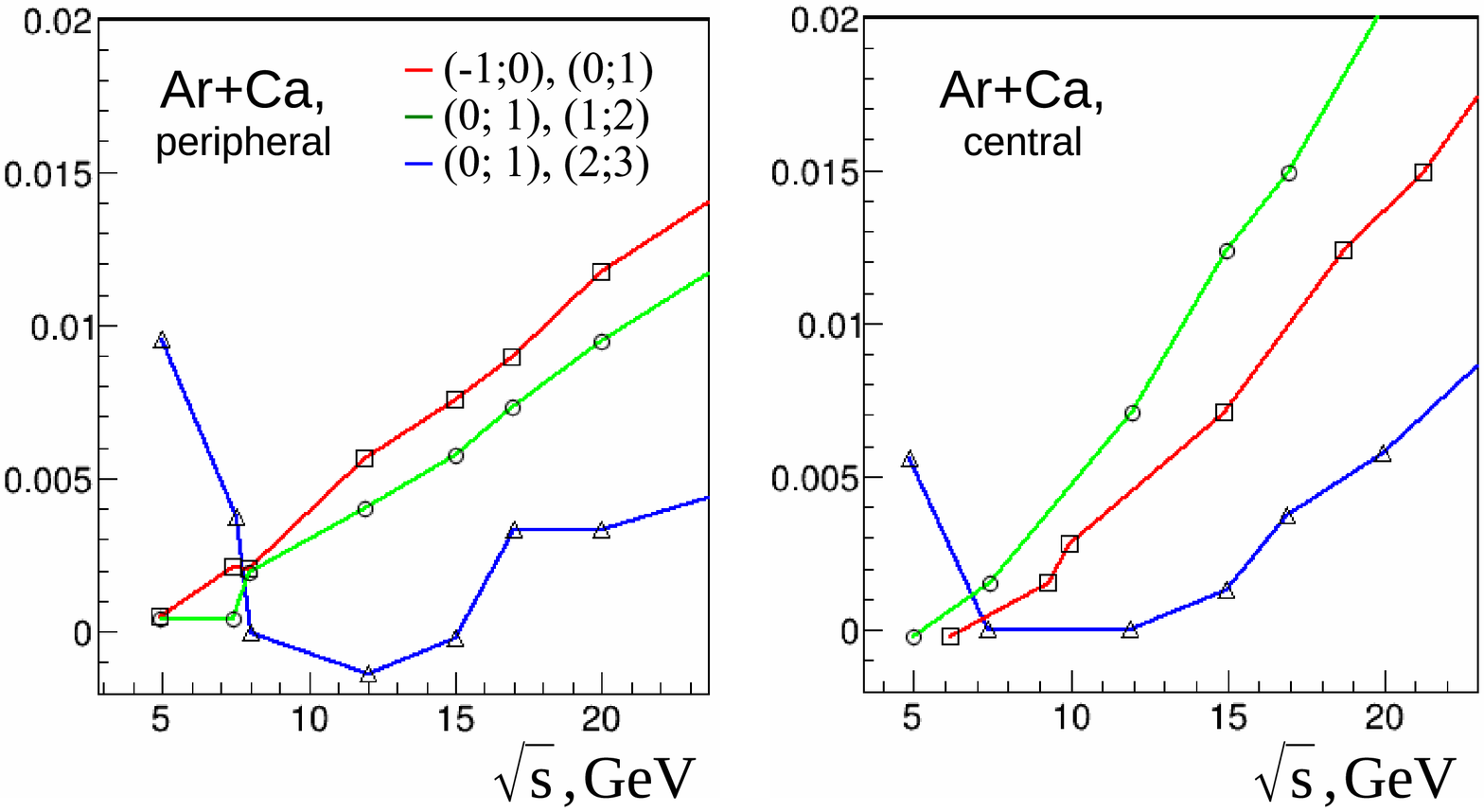}
          \put(3,54){ \Large{$\mathrm{b}_{\mathrm{p_t-p_t}}$}}
\end{overpic}
\caption{Dependence of pt-pt correlation coefficient in three rapidity windows configurations on the collision energy in Ar+Ca collisions.}
\label{fig8}
\end{figure*}

In the results of the calculation of $b_{\mathrm{pt-pt}}$ one observes a monotonic increase  with the collision energy mostly.
However, in Ar+Ca there is a non-monotonic dependence with a minimum around $\sqrt{s}$=17 GeV. One should stress that the pt-pt correlation coefficient behaves non-monotonically only in case of middle-forward rapidity configurations, while in the other two cases (midrapidity ones) the dependence is smooth. This picture suggests the existence of a phase transition in the area of forward rapidity (note that the larger rapidity corresponds to higher baryon chemical potential) with smooth crossover in midrapidity (where the baryon density is minimal).
Note that similar non-monotonic behavior has been obtained experimentally
in the energy range 10-20 GeV for different fluctuational observables
(for example, the deviations from Poisson of net-proton fluctuations at RHIC \cite{NPRHIC}), and it was attributed to the existence of the phase
transition and the critical point 
on the QCD phase diagram.
In the Fig. \ref{fig8} we show a more 
detailed study of pt-pt correlations in Ar+Ca collisions in the energy range between 5 and 20 GeV. 

We performed these addition calculations in order to study the shape of the observed dip in pt-pt correlation 
energy dependence. The results confirmed previous calculation. The central position of the dip in peripheral
collisions is found at higher energy than in 
central, which suggests that the phase transition happens earlier in more central collisions.
Overall, the obtained picture advises, that being applied to the
range of non-zero baryon chemical potential, the present model, based on the string fusion approach, reproduces the phase transition one expects 
in the QCD equation of state.

The calculations  highlight the importance of Ar+Ca run in the beam  energy  and system size scan by the NA61 Collaboration at the SPS scheduled on 2015 year \cite{ArCa}. The experimental study of the long-range pt-pt correlations, as well as different strongly intensive observables \cite{SI0,SI1}, would be crucial  for the improvement of our understanding of the phase diagram of the strongly interacting matter.

\pagebreak

\section{Conclusions}
String fusion approach to the quark-gluon plasma formation 
at non-zero baryon chemical potential has been proposed.
A model for the string fusion, accounting finite rapidity width of strings, for pp, pA and AA collisions is developed and applied at the SPS energies.
Long-range correlation functions and coefficients are studied.
Smooth monotonic behavior of n-n and pt-n correlation with energy
with non-monotonic pt-pt correlation dependence in Ar+Ca collisions have been obtained.
\newline
A more detailed scan, including calculation of correlations in narrow centrality classes, is required, as well as
an extension of the model for net-charge and net-proton fluctuation and correlation studies, exploring the strongly intensive variables.

\vspace{0.2cm}

\section{Acknowledgements}
\scalebox{.945}[1.0]{
V. Kovalenko acknowledges Saint-Petersburg State University for the research grant 11.38.193.2014} \scalebox{.95}[1.0]{and Special Rector's Scholarship. He is also grateful to the Dynasty Foundation.}


\vspace{0.2cm}

\begin{thebibliography}{99}

\bibitem{PD2} Y.~Aoki, G.~Endrodi, Z.~Fodor, S.~D.~Katz, and K.~K.~Szabo, \emph{The order of the quantum chromodynamics transition predicted by the standard model of particle physics}, Nature \textbf{443}, 675 (2006),  \href{http://arxiv.org/abs/hep-lat/0611014}{\tt hep-lat/0611014}.
\bibitem{HW}C. Blume, \emph{Studies on the QCD Phase Diagram at SPS and FAIR}, J. Phys. Conf. Ser. \textbf{422}, 012022 (2013).
\bibitem{HW1}   C.~Schmidt,   \emph{Lattice QCD at finite density},   \pos{PoS(LAT2006)021} (2006),   \href{http://arxiv.org/abs/hep-lat/0610116}{\tt hep-lat/0610116}.  
\bibitem{HW2} O. Philipsen, \emph{Lattice calculations at non zero chemical potential}, \pos{PoS(ConfinementVIII)0111} (2009).
\bibitem{PD0} F.~R.~Brown, et al., \emph{On the existence of a phase transition for QCD with three light quarks}, Phys. Rev. Lett. \textbf{65}, 2491 (1990).
\bibitem{PD1} A.~A.~Khan, et al., \emph{Phase structure and critical temperature of two-flavor QCD with a renormalization group improved gauge action and clover improved Wilson quark action}, Phys. Rev. D \textbf{63}, 034502 (2001), \href{http://arxiv.org/abs/hep-lat/0008011}{\tt hep-lat/0008011}.
\bibitem{PD3}  Z.~Fodor, S.~D.~Katz, \emph{Critical point of QCD at finite T and $\mu$, lattice results for physical quark masses}, JHEP \textbf{0404}, 050 (2004), \href{http://arxiv.org/abs/hep-lat/0402006}{\tt hep-lat/0402006}.
\bibitem{HWn} P. de Forcrand and O. Philipsen, \emph{The chiral critical line of N${_f}$ = 2+1 QCD at zero and non-zero baryon density}, JHEP 01, 077 (2007), \href{http://arxiv.org/abs/hep-lat/0607017}{\tt hep-lat/0607017}.
\bibitem{HWm} O. Philipsen, \emph{Status of the QCD Phase Diagram from Lattice Calculations}, Acta Phys. Polon. Supp. \textbf{5}, 825-835 (2012), \href{http://arxiv.org/abs/1111.5370}{\tt arXiv:1111.5370 [hep-ph]}.
\bibitem{PDE0}   M.~Gazdzicki,   \emph{Onset of Deconfinement and Critical Point: NA49 and NA61/SHINE at the CERN SPS},   Eur.\ Phys.\ J.\ ST {\bf 155}, 37 (2008), \href{http://arxiv.org/abs/0801.4919}{\tt 	arXiv:0801.4919 [nucl-ex]}.
\bibitem{PDE1} C. Hohne, \emph{Physics of compressed baryonic matter}, J. Phys. Conf. Ser. \textbf{420}, 012016 (2013).
\bibitem{BES0}   M.~Gazdzicki (NA49 and NA61/SHINE Collaborations),   \emph{NA49/NA61: results and plans on beam energy and system size scan at the CERN SPS},   J.\ Phys.\ G {\bf 38}, 124024 (2011), \href{http://arxiv.org/abs/1107.2345}{\tt 	arXiv:1107.2345 [nucl-ex]}.   
\bibitem{BES1} G.~Odyniec,  \emph{The RHIC Beam Energy Scan program in STAR and what's next ...}, J. Phys. Conf. Ser. \textbf{455}, 012037 (2013).
\bibitem{BES2} P. Staszel,  \emph{CBM experiment at FAIR,}  Acta Phys. Pol. B \textbf{41}, 341 (2010).
\bibitem{BES3}   V.~Toneev,   \emph{The NICA/MPD project at JINR (Dubna)},    \pos{PoS(CPOD 07)057} (2007),   \href{http://arxiv.org/abs/0709.1459}{\tt 	arXiv:0709.1459 [nucl-ex]}.   
\bibitem{PDO0} M.~S.~Borysova, \emph{Quark-gluon plasma signals in CBM experiment}, J. Phys. Studies \textbf{14}, 3203 (2010).
\bibitem{NA61}N.~Antoniou, et al. (NA61 Collaboration). \emph{Study of hadron production in hadron-nucleus and nucleus-nucleus collisions at the CERN SPS}. \href{http://cds.cern.ch/record/995681}{SPSC-P-330, CERN-SPSC-2006-034} (2006). 

\bibitem{SF1}M.~A.~Braun, C.~Pajares, J.~Ranft,  \emph{Fusion of strings vs. percolation and the transition to the quark-gluon plasma}, Int.~J.~Mod.~Phys.~A {\bfseries 14} 2689 (1999).   
\bibitem{SF2} N.~S.~Amelin, M.~A.~Braun, C.~Pajares, \emph{Multiple production in the Monte Carlo string fusion model}, {Phys. Lett. B} \textbf{306}, 312 (1993). 
\bibitem{F33} M. A. Braun and C. Pajares, \emph{Particle production in nuclear collisions and string interactions}, Phys. Lett. B \textbf{287}, 154 (1992).
\bibitem{F33a} M. A. Braun and C. Pajares, \emph{A Probabilistic model of interacting strings}, Nucl. Phys. B \textbf{390}, 542 (1993).
\bibitem{OL1} N. S. Amelin, N. Armesto, M.A. Braun, E. G. Ferreiro, and C. Pajares, \emph{Long and short range correlations and the search of the quark gluon plasma}, Phys. Rev. Lett. \textbf{73}, 2813 (1994).
\bibitem{TP1} O.~Kochebina, G.~Feofilov,  \emph{Onset of "ridge phenomenon" in AA and pp collisions and percolation string model}, in {Proc. XX Baldin ISHEPP} (2010), \href{http://arxiv.org/abs/1012.0173}{\tt 	arXiv:1012.0173 [hep-ph]}.
\bibitem{TP2} M.~A.~Braun, C.~Pajares, V.~V.~Vechernin, \emph{Anisotropic flows from colour strings: Monte Carlo simulations}, {Nucl. Phys. A} \textbf{906}, 14 (2013),  	\href{http://arxiv.org/abs/1204.5829}{\tt 	arXiv:1204.5829 [hep-ph]}.
\bibitem{RP1} R.~P.~Scharenberg, B.~K.~Srivastava, A.~S.~Hirsch,  \emph{Percolation of color sources and the equation of state of QGP in central Au-Au collisions at $\sqrt{s_{NN}}$= 200 GeV},  Eur.~Phys.~J.~C \textbf{71}, 1510 (2011),  \href{http://arxiv.org/abs/1006.3260}{\tt 	arXiv:1006.3260 [nucl-ex]}.
\bibitem{RP2} J.~Dias de Deus, A.~S.~Hirsch, C.~Pajares, R.~P.~Scharenberg, B.~K.~Srivastava,  \emph{Percolation of Color Sources and the Shear Viscosity of the QGP in Central A-A Collisions at RHIC and LHC Energies},  \href{http://arxiv.org/abs/1106.4271}{\tt 	arXiv:1106.4271 [nucl-ex]} (2011).
\bibitem{RP3}   B.~K.~Srivastava,   \emph{Percolation Approach to Initial Stage Effects in High Energy Collisions},   Nucl.\ Phys.\ A {\bf 926}, 142 (2014),   \href{http://arxiv.org/abs/1402.2306}{\tt 	arXiv:1402.2306 [nucl-ex]}.   
\bibitem{RP4}   J.~Dias de Deus, A.~S.~Hirsch, C.~Pajares, R.~P.~Scharenberg, and B.~K.~Srivastava,   \emph{Clustering of color sources and the shear viscosity of the QGP in heavy ion collisions at RHIC and LHC energies},   Eur.\ Phys.\ J.\ C {\bf 72}, 2123 (2012).   
\bibitem{OL2} V. N. Kovalenko, \emph{Modeling of exclusive parton distributions and long-range rapidity correlations in proton-proton collisions at the LHC energies}, Phys. Atom. Nucl. \textbf{76}, 1189 (2013), \href{http://arxiv.org/abs/1211.6209}{\tt 	arXiv:1211.6209 [hep-ph]}.
\bibitem{YF1PoS2} V. Kovalenko, V. Vechernin,  \emph{Model of pp and AA collisions for the description of long-range correlations},  \pos{PoS(Baldin ISHEPP XXI)077} (2012).
\bibitem{WoSa} H. De Vries, C. W. De Jager, and C. De Vries, Atom. Data Nucl. Data Tabl. \textbf{36}, 495 (1987). 
\bibitem{OL7} C. Flensburg, G. Gustafson, and L. Lonnblad, Eur. Phys. J. C \textbf{60}, 233 (2009).
\bibitem{OL8} G. Gustafson, Acta Phys. Polon. B \textbf{40}, 1981 (2009). 
\bibitem{OL6} V. Kovalenko,  \emph{Monte Carlo model for pp, pA and AA collisions at high energy: parameters tuning and results}, \pos{PoS(QFTHEP 2013)052} (2013). 
\bibitem{VK1}    V.~V.~Vechernin and R.~S.~Kolevatov,   \emph{Long-range correlations between transverse momenta of charged particles produced in relativistic nucleus-nucleus collisions},   Phys.\ Atom.\ Nucl.\  {\bf 70}, 1809 (2007)   [Yad.\ Fiz.\  {\bf 70}, 1858 (2007)].   
\bibitem{VK2}    V.~V.~Vechernin and R.~S.~Kolevatov,   \emph{On multiplicity and transverse-momentum correlations in collisions of ultrarelativistic ions},   Phys.\ Atom.\ Nucl.\  {\bf 70}, 1797 (2007)   [Yad.\ Fiz.\  {\bf 70}, 1846 (2007)].   
\bibitem{VSM}   J.~Alvarez-Muniz, R.~Conceicao, J.~Dias de Deus, M.~C.~Espirito Santo, J.~G.~Milhano and M.~Pimenta,   \emph{A Model for net-baryon rapidity distribution},   Eur.\ Phys.\ J.\ C {\bf 61} 391 (2009),   \href{http://arxiv.org/abs/0903.0957}{\tt 	arXiv:0903.0957 [hep-ph]}.   
\bibitem{Adam:2015mya}    J.~Adam et al. (ALICE Collaboration),   \emph{Forward-backward multiplicity correlations in pp collisions at $\sqrt{s}$=0.9, 2.76 and 7 TeV},   \href{http://arxiv.org/abs/1502.00230}{\tt  arXiv:1502.00230 [nucl-ex]} (2015).   
\bibitem{TbD0} A. Capella and J. Tran Thanh Van, ``Long Range Rapidity Correlations in Hadron - Nucleus Interactions,'' Phys. Rev. \textbf{D} 29, 2512 (1984).
\bibitem{TbD1} T. Alexopoulos et al. (E735 Collaboration), \emph{Charged particle multiplicity correlations in $\mathrm{p\bar{p}}$ collisions at $\sqrt{s}$ = 0.3 TeV to 1.8 TeV}, Phys. Lett. B \textbf{353}, 155-160 (1995).
\bibitem{TbD2} K. Alpgard et al. (UA5 Collaboration), \emph{Forward-Backward Multiplicity Correlations in p anti-p Collisions at 540-GeV}, Phys. Lett. B \textbf{123}, 361 (1983).
\bibitem{TbD3} G. Aad et al. (ATLAS Collaboration), \emph{Forward-backward correlations and charged-particle azimuthal distributions in pp interactions using the ATLAS detector}, JHEP \textbf{1207}, 019 (2012), \href{http://arxiv.org/abs/1203.3100}{\tt 	arXiv:1203.3100 [hep-ex]}.
\bibitem{NA49baldin}  C. Alt et al. (NA49 Collaboration) and G. A. Feofilov et al. (SPbSU group), \emph{Long-range correlations in PbPb collisions at 158 A*GeV}, in Proc. Relativistic Nuclear Physics and Quantum Chromodynamics, Ed. by A. N. Sissakian, V. V. Burov, and A. I. Malakhov (JINR, Dubna, 2005), Vol. 1, p. 222.
\bibitem{OL15} V. Kovalenko and V. Vechernin,  \emph{Long-range rapidity correlations in high energy AA collisions in Monte Carlo model with string fusion}, Eur. Phys. J. Web of Conf. \textbf{66}, 04015 (2014), \href{http://arxiv.org/abs/1308.6618}{\tt 	arXiv:1308.6618 [nucl-th]}. 
\bibitem{OL25}   V.~Kovalenko and V.~Vechernin,   \emph{Forward-backward multiplicity correlations in pp collisions at high energy in Monte Carlo model with string fusion},  \href{http://arxiv.org/abs/1410.3884}{\tt 	arXiv:1410.3884 [hep-ph]} (2014).   
\bibitem{VF}    V.~V.~Vechernin,   \emph{Correlations between multiplicities in rapidity and azimuthally separated windows},   \href{http://arxiv.org/abs/1210.7588}{\tt 	arXiv:1210.7588 [hep-ph]} (2012).   
\bibitem{Andronov}  E.~Andronov, V.~Vechernin,  \emph{The correlation between transverse momentum and multiplicity of charged particles in a two-component model}, \pos{PoS(QFTHEP 2013)054} (2014). 
\bibitem{NPRHIC}    L.~Adamczyk, et al. (STAR Collaboration),   \emph{Beam energy dependence of moments of the net-charge multiplicity distributions in Au+Au collisions at RHIC},   Phys.\ Rev.\ Lett.\  {\bf 113}, 092301 (2014), \href{http://arxiv.org/abs/1402.1558}{\tt 	arXiv:1402.1558 [nucl-ex]}.   
\bibitem{ArCa}    D.~Kuchler, M.~O`Neil, R.~Scrivens and R.~Thomae,   \emph{Preparation of a primary argon beam for the CERN fixed target physics},   Rev.\ Sci.\ Instrum.\  {\bf 85}, 02A954 (2014).   
\bibitem{SI0}   M.~Gazdzicki, M.~I.~Gorenstein and M.~Mackowiak-Pawlowska,  \emph{Normalization of strongly intensive quantities},   Phys.\ Rev.\ C {\bf 88}, 024907 (2013), \href{http://arxiv.org/abs/1303.0871}{\tt 	arXiv:1303.0871 [nucl-th]}.   
\bibitem{SI1}   M.~I.~Gorenstein and K.~Grebieszkow,   \emph{Strongly Intensive Measures for Transverse Momentum and Particle Number Fluctuations},   Phys.\ Rev.\ C {\bf 89}, 034903 (2014), \href{http://arxiv.org/abs/1309.7878}{\tt 	arXiv:1309.7878 [nucl-th]}.   


\end{thebibliography}
\end{document}